\documentclass[journal,fleqn]{IEEEtran}
\usepackage{amsmath,amsfonts}
\usepackage{amsthm}
\newtheorem{proposition}{Proposition}
\usepackage{url}
\usepackage{cite}
\usepackage{lipsum}
\usepackage{bm}
\usepackage[colorlinks=true,
            linkcolor=blue,      
            citecolor=blue,      
            urlcolor=blue]{hyperref}
\usepackage{multirow}
\usepackage{graphicx}
\usepackage{tabularx}
\usepackage{booktabs}
\usepackage{algorithm}
\usepackage{algpseudocode}
\usepackage{array}
\newcolumntype{Y}{>{\centering\arraybackslash}X}
\newcommand{\refig}[1]{Fig. \ref{#1}}
\newcommand{\mt}[1]{\textit{#1}} 
\usepackage{enumitem}
\newlist{Indices}{description}{1}
\setlist[Indices]{style=nextline, leftmargin=10mm, labelindent=0cm, itemsep=0.2mm, font=\normalfont}

\newlist{Parameters}{description}{1}
\setlist[Parameters]{style=nextline, leftmargin=18mm, labelindent=0cm, itemsep=0.2mm, font=\normalfont}

\newlist{Variables}{description}{1}
\setlist[Variables]{style=nextline, leftmargin=15mm, labelindent=0cm, itemsep=0.2mm, font=\normalfont}

\begin{document}

\title{\huge \vspace*{-5mm} A Neural Column-and-Constraint Generation Method for Solving Two-Stage Stochastic Unit Commitment}

\author{Zhentong Shao, \textit{Member, IEEE}, Jingtao Qin, \textit{Graduate Student Member, IEEE}, Nanpeng Yu, \textit{IEEE Senior Member}

\thanks{Z. Shao, J. Qin and N. Yu are with the Department of Electrical and Computer Engineering, University of California, Riverside, CA 92521 USA (e-mail: zhentons@ucr.edu; jqin020@ucr.edu; nyu@ece.ucr.edu).}

\vspace{-9mm}
}

\maketitle

\begin{abstract}
    Two-stage stochastic unit commitment (2S-SUC) problems have been widely adopted to manage the uncertainties introduced by high penetrations of intermittent renewable energy resources. While decomposition-based algorithms such as column-and-constraint generation has been proposed to solve these problems, they remain computationally prohibitive for large-scale, real-time applications. In this paper, we introduce a Neural Column-and-Constraint Generation (Neural CCG) method to significantly accelerate the solution of 2S-SUC problems. The proposed approach integrates a neural network that approximates the second-stage recourse problem by learning from high-level features of operational scenarios and the first-stage commitment decisions. This neural estimator is embedded within the CCG framework, replacing repeated subproblem solving with rapid neural evaluations. We validate the effectiveness of the proposed method on the IEEE 118-bus system. Compared to the original CCG and a state-of-the-art commercial solver, Neural CCG achieves up to 130.1$\times$ speedup while maintaining a mean optimality gap below 0.096\%, demonstrating its strong potential for scalable stochastic optimization in power system.
\end{abstract}

\vspace{-1mm}

\vspace*{-3mm}
\section{Introduction}
\IEEEPARstart{T}{he} growing penetration of uncertain renewable energy sources (RES) introduces significant challenges to the reliable and efficient daily operation of modern power systems. To manage these uncertainties, the two-stage stochastic unit commitment (2S-SUC) model has emerged as a foundational framework in operations planning \cite{wu2007stochastic}. A widely adopted solution approach involves representing uncertainty via a finite set of scenarios, thereby formulating the problem as a deterministic extensive-form model amenable to commercial solvers. However, this method often incur substantial computational overhead, especially when the scenario set is large and the recourse stage involves nonlinear optimization \cite{hu2015robust}.

To address the computational challenges associated with extensive-form formulations, decomposition-based algorithms such as Benders decomposition \cite{shao2023risk} and column-and-constraint generation (CCG) \cite{zeng2013solving} have been widely adopted. Among these, CCG has shown superior computational efficiency and is often favored for large-scale stochastic optimization problems. Empirical results indicate that CCG can deliver a 3 to 6$\times$ speed up over solving the extensive form directly. However, our analysis reveals that a significant portion of the total computation time -- approximately 60\% -- is spent solving recourse problems within the CCG framework. This proportion can escalate to over 95\% when the number of scenarios exceeds 1,000, highlighting a key bottleneck in scalability.

Motivated by this observation, we propose a neural CCG approach to accelerate the solution process. In this framework, the second-stage recourse problem is approximated using a deep neural network that learns a mapping from uncertainty features and commitment decisions to the corresponding recourse costs. By replacing the exact recourse problem with this learned surrogate, the proposed method significantly alleviates the computational burden associated with traditional CCG. Simulation studies on the IEEE 118-bus system demonstrate that the neural CCG method substantially improves solution speed while preserving high solution quality.

\vspace*{-1mm}
\section{Problem Formulation}
\vspace*{-1mm}
\subsection{Unit Commitment Model}
\vspace*{-1mm}
This paper investigates the unit commitment (UC) problem incorporating transmission constraints modeled using power transfer distribution factor (PTDF). The corresponding mathematical formulation is presented below.
\begin{align}
    &\min_{p_{g,t}, z_{g,t}, u_{g,t}, v_{g,t}} \sum_{t \in \mathcal{N}_{t}} \! \sum_{g \in \mathcal{N}_{g}}  \! \left( C_{g}(p_{g,t}) + C^{\mt{SU}}_{g} u_{g,t} + C^{\mt{SD}}_{g} v_{g,t} \right)  \!  \! \label{uc:obj}
    \\
    &\text{s.t. } \sum_{g \in \mathcal{N}_{g}} \nolimits p_{g,t} - \sum_{n \in \mathcal{N}_{n}} \nolimits d_{n,t} = 0, \hspace{15mm} \forall t \in \mathcal{N}_{t} \label{uc:balance}
    \\
    &-\overline{F}_{l} \leq \sum_{g \in \mathcal{N}_{g}} \nolimits \Gamma^{G}_{l,g} p_{g,t} - \sum_{n \in \mathcal{N}_{n}} \nolimits \Gamma^{D}_{l,n} d_{n,t} \leq \overline{F}_{l}, \nonumber
    \\
    & \hspace{56mm} \forall l \in \mathcal{N}_{l},~\forall t \in \mathcal{N}_{t} \label{uc:transmission}
    \\
    &z_{g,t} \underline{P}_g \leq p_{g,t} \leq z_{g,t} \overline{P}_g,  \hspace{19.5mm} \forall g \in \mathcal{N}_g,~\forall t \in \mathcal{N}_{t} \label{uc:capacity}
    \\
    &- \! R_g^{\mt{DN}} \leq p_{g,t} - p_{g,t-1} \leq R_g^{\mt{UP}}, \hspace{9.5mm} \forall g \in \mathcal{N}_g,~ \forall t \in \mathcal{N}_{t}\label{uc:ramping}
    \\
    &u_{g,t} - v_{g,t} = z_{g,t} - z_{g,t-1}, \hspace{15.5mm} \forall g \in \mathcal{N}_g,~\forall t \in \mathcal{N}_{t} \label{uc:logic}
    \\
    &\sum_{\tau = t - T_g^{\mt{UP}} + 1}^{t} \!\! u_{g,\tau} \leq z_{g,t}, \hspace{3.5mm} \forall g \in \mathcal{N}_g,~\forall t \in \{T_g^{\mt{UP}}, \ldots, |\mathcal{N}_{t}|\} \label{uc:up_time}
    \\
    &\sum_{\tau = t - T_g^{\mt{DN}} + 1}^{t} \!\!\! v_{g,\tau} \leq 1 \!-\! z_{g,t},\hspace{0.5mm}\forall g \in \mathcal{N}_g, \hspace{0.5mm}\forall t \in \{T_g^{\mt{DN}}, \ldots, |\mathcal{N}_{t}|\} \!\! \label{uc:down_time}
\end{align}
where the objective \eqref{uc:obj} minimizes total generation, start-up, and shut-down costs. Binary variables $z_{g,t}$, $u_{g,t}$, and $v_{g,t}$ denote generator commitment, start-up, and shut-down status, respectively. $p_{g,t}$ represents generator output. Constraint \eqref{uc:balance} enforces power balance. Constraint \eqref{uc:transmission} limits line flows using PTDF. Constraints \eqref{uc:capacity} and \eqref{uc:ramping} impose capacity and ramping limits. Constraint \eqref{uc:logic} encodes logical conditions, and \eqref{uc:up_time}–\eqref{uc:down_time} enforce minimum up and down times. Detailed model specifications can be found in \cite{shao2023linear}.

\vspace*{-2mm}
\subsection{2S-SUC Formulation}
\vspace*{-1mm}
To account for uncertainties in renewable generation and load demand, the UC model \eqref{uc:obj}–\eqref{uc:down_time} is reformulated into a 2S-SUC model. In this formulation, the net load (i.e., demand minus renewable generation) is treated as the source of uncertainty. The resulting extensive-form of the 2S-SUC model is presented below in a compact form. 

\vspace*{-5mm}
\begin{align}
    &\min_{\bm{x}, \bm{z}}~ \bm{c}^{\mathsf{T}} \bm{x} + \sum_{s \in \mathcal{S}} \nolimits \pi_s \, Q(\bm{z}, \bm{\xi}_{s})
    \\
    \text{s.t.}~ &\bm{x} \in \mathcal{X}(\bm{z}),
    \\
    \vspace{-1mm}
    &Q(\bm{z}, \bm{\xi}_{s}) = 
    \left\{
    \begin{aligned}
        \min_{\bm{y}_{s}} \quad & \bm{q}^{\mathsf{T}} \bm{y}_{s} \\
        \text{s.t.} \quad & \bm{W} \bm{y}_{s} \geq \bm{h}(\bm{\xi}_{s})
        , \quad \bm{y}_{s} \in \mathcal{Y}(\bm{z})
    \end{aligned}
    \right\}
\end{align}
where $\mathcal{S} = \{\bm{\xi}_{1}, \ldots, \bm{\xi}_{S} \}$ denotes a set of $S$ scenarios, with associated probabilities $\pi_{s}$ for each $s \in \mathcal{S}$. The vector $\bm{x}$ comprises the variables $u_{g,t}$ and $v_{g,t}$, while $\bm{z}$ represents the vector of $z_{g,t}$. The vector $\bm{y}_{s}$ corresponds to the generator outputs under the $s$-th scenario. The function $Q(\bm{z}, \bm{\xi})$ denotes the recourse problem.

\section{Neural Column-and-Constraint Generation}

\subsection{The Classical CCG Method}
\vspace{-1mm}
The classical CCG method for solving the 2S-SUC problem is outlined in Algorithm~\ref{alg:org-ccg}, where the master problem and subproblem are solved iteratively until convergence. A key assumption in the standard CCG framework is the feasibility of the subproblem in each iteration. In practice, this assumption is often ensured by incorporating penalty terms into the objective function and introducing slack variables, such as load shedding, to accommodate potential infeasibilities.

\vspace*{-2mm}
\begin{algorithm}[H]
\caption{The Classical CCG Method for 2S-SUC}\label{alg:org-ccg}
\begin{algorithmic}[1]
\State Initialize iteration counter $k \leftarrow 0$, tolerance $\varepsilon$, set $\mathcal{S}^{k} = \emptyset$.
\State Set initial bounds $\mt{LB} \leftarrow -\infty$, $\mt{UB} \leftarrow +\infty$.

\State \textbf{Repeat}

\State \quad \parbox[t]{79mm}{Solve the master problem (\textbf{MP}):
\vspace{-1mm}
\begin{gather}
\begin{aligned}
    \quad \quad \min_{\bm{x},\bm{z},\eta} ~~& \bm{c}^{\mathsf{T}}\bm{x} + \sum_{s\in \mathcal{S}^{k}} \nolimits \pi_{s}\eta_{s}
    \\
    \quad \quad \text{s.t.}~~& \bm{x}\in\mathcal{X}(\bm{z}),~ \eta_{s} \geq Q(\bm{z}, \bm{\xi}_{s}),~\forall s \in \mathcal{S}^{k}.
\end{aligned}
\label{eq:MP}
\end{gather}}

\State \quad \parbox[t]{79mm}{Get the optimal solution $(\bm{x}^{k},\bm{z}^{k})$ \& objective value $\theta^k$,  Set $\mt{LB} \leftarrow \theta^k$.}

\vspace{1.5mm}

\State \quad \parbox[t]{79mm}{Solve subproblem (\textbf{SP}) for all $s \in \mathcal{S}$:
\vspace{-1mm}
\begin{gather*}
    ~~~~~~\min_{\bm{y}_s} \left\{ \bm{q}^{\mathsf{T}} \bm{y}_s \mid \bm{W} \bm{y}_s \geq \bm{h}(\bm{\xi}_s),~ \bm{y}_s \in \mathcal{Y}(\bm{z}^k) \right\}
\end{gather*}}

\State \quad Compute $\mt{UB}^k = \bm{c}^{\mathsf{T}}\bm{x}^{k} + \sum_{s\in \mathcal{S}} \pi_{s} Q(\bm{z}^{k}, \bm{\xi}_{s})$.

\vspace{1mm}

\State \quad Update upper bound: $\mt{UB} \leftarrow \min(\mt{UB}, \mt{UB}^k)$.

\vspace{1mm}

\State \quad \parbox[t]{79mm}{Select the most violating or highest-cost scenario(s) $\bm{\xi}_{s}^{*}$ to be added to the scenario set:
\vspace{-1mm}
\[
\quad \quad \mathcal{S}^{k+1}\leftarrow \mathcal{S}^{k}\cup\{\bm{\xi}_{s}^{*}\}.
\]
}

\State \quad Increment the iteration counter $k \leftarrow k + 1$.

\State \quad Update the optimality gap: $\text{Gap} = \mt{UB}-\mt{LB}$.

\State \textbf{Until} $\text{Gap} \leq \varepsilon$

\State Return the optimal solution $(\bm{x}^{*},\bm{z}^{*})$ \& objective value $\theta^{*}$.
\end{algorithmic}
\end{algorithm}

\vspace*{-2mm}

\vspace*{-5mm}
\subsection{The Proposed Neural CCG Method}
\begin{figure}[tbp]
    \vspace{-3mm}
    \centering
    \includegraphics[width=0.8\linewidth]{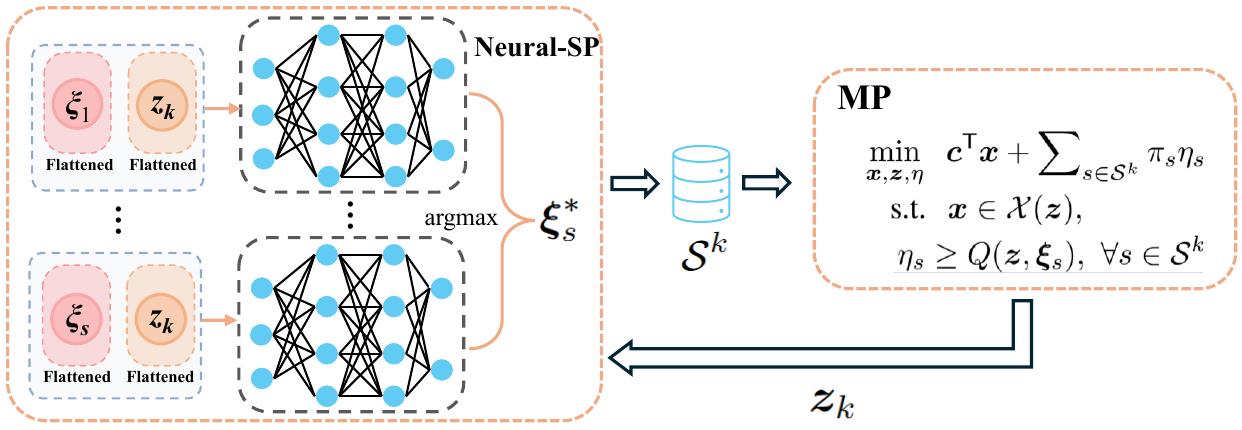}
    \vspace{-2mm}
    \caption{The overall neural CCG framework.}
    \vspace{-2mm}
    \label{fig:nn-structure}
\end{figure}

The proposed neural CCG method is outlined in Algorithm~\ref{alg:neu-ccg} with the overall framework illustrated in \refig{fig:nn-structure}. In this approach, the resource problem is approximated by a neural network (NN), which is formulated as follows:

\vspace{-5mm}
\begin{align}
    Q(\bm{z}, \bm{\xi}_s) \approx \mt{NN} \left(\bm{z}, \bm{\xi}_{s} \right).
\end{align}
\vspace{-5mm}

Here, $\mt{NN} \left(\bm{z}, \bm{\xi}_{s} \right)$ denotes the trained neural network that approximates the mapping $\left(\bm{z}, \bm{\xi}_{s} \right) \rightarrow Q$. Since the neural network outputs a scalar value, high prediction accuracy can be achieved using a simple and lightweight architecture. Notably, the \textbf{Recourse Action} step in Algorithm~\ref{alg:neu-ccg} requires only a forward pass through the neural network, which can be executed in a matter of seconds, significantly accelerating the overall CCG solution process.

\begin{algorithm}[H]
\caption{Neural CCG Method for 2S-SUC}\label{alg:neu-ccg}
\begin{algorithmic}[1]
\State Initialize iteration counter $k \leftarrow 0$, tolerance $\varepsilon$, set $\mathcal{S}^{k} = \emptyset$.

\State \textbf{Repeat}

\State \quad \parbox[t]{79mm}{Solve the \textbf{MP} defined in \eqref{eq:MP}.}

\State \quad \parbox[t]{79mm}{Get the optimal solution $(\bm{x}^{k},\bm{z}^{k})$ \& objective value $\theta^k$}

\vspace{1.5mm}

\State \quad \parbox[t]{79mm}{\textbf{Recourse Action}: Select the most violating or highest-cost scenario(s) $\bm{\xi}_{s}^{*}$ by
\begin{align*}
    \quad \quad \bm{\xi}_{s}^{*} = \text{arg} \max_{s \in \mathcal{S}} \mt{NN}(\bm{z}_{k}, \bm{\xi}_{s}).
\end{align*}
}

\State \quad \parbox[t]{79mm}{Add the selected scenario to the scenario set:
\[
\quad \quad \mathcal{S}^{k+1}\leftarrow \mathcal{S}^{k}\cup\{\bm{\xi}_{s}^{*}\}.
\]
}

\State \quad Increment the iteration counter $k \leftarrow k + 1$.

\vspace{1mm}

\State \textbf{Until} the following criteria holds:
\vspace{-1.5mm}
\begin{align}
    \quad \quad \quad \max_{s \in \mathcal{S}} \mt{NN}(\bm{z}_{k}, \bm{\xi}_{s}) \leq \max_{s \in \mathcal{S}^{k}} \mt{NN}(\bm{z}_{k}, \bm{\xi}_{s}) + \varepsilon. \label{eq:terminate}
\end{align}

\vspace{-0.5mm}
\State Return the optimal solution $(\bm{x}^{*},\bm{z}^{*})$ \& objective value $\theta^{*}$.
\end{algorithmic}
\end{algorithm}

\vspace{-1mm}
\noindent \textbf{Convergence}:
Since our method deviates from the standard CCG procedure, the convergence guarantee of the  classical algorithm no longer apply. To address this, we present the following proposition, with its proof provided in Appendix~\ref{appendix-proof}.

\begin{proposition}
If $\bm{z} \in \mathcal{Z}$ is finite, the neural CCG algorithm terminates after a finite number of iterations.
\end{proposition}

\subsection{Architecture of Neural Network}
The recourse problem is approximated using a multilayer perceptron (MLP) composed of fully connected layers with ReLU activation functions, as illustrated in \refig{fig:nn-structure}. The scenario vector $\bm{\xi}_{s}$ and the commitment decision vector $\bm{z}$ are flattened and concatenated to form the input to the neural network. A simple and lightweight MLP has proven effective in supporting the neural CCG framework. Although more advanced architectures, such as graph neural networks (GNNs) \cite{qin2023solve}, may offer improved feature extraction and performance, their investigation is beyond the scope of this letter and is left for future work.

Ensuring the feasibility of the recourse problem is a critical aspect of the neural CCG framework. To this end, penalty terms are incorporated into the objective function to relax key constraints, particularly those related to load balancing and transmission limits. While these penalties are often chosen heuristically, typically as multiples of electricity prices, our results show that setting them equal to the Lagrange multipliers associated with the corresponding constraints in the UC model under the mean or expected scenario yields significantly improved performance. This strategy enhances both the solution accuracy and the overall effectiveness of the neural CCG approach.

\vspace*{-1mm}
\section{Simulation Results}
\subsection{Test System and Parameter Settings}
The proposed Neural CCG method is evaluated on a modified IEEE 118-bus system \cite{zimmerman2010matpower}. All transmission line capacities are uniformly set to 350 MW, and generator ramping limits are constrained to 25\% of their respective capacities. Net load profiles are randomly generated with variability ranging from 70\% to 100\% of the original value. Performance is assessed using two key metrics: average optimality gap and speedup. The optimality gap is measured relative to the the benchmark solution obtained using Gurobi and is averaged over $K = 100$ test instances. Specifically, the gap is computed with respect to $\mt{obj}_{\text{true}}$, the objective value from directly solving the problem via Gurobi. Speedup is defined as the ratio between Gurobi’s runtime and that of the proposed neural CCG method.

The neural network architecture consists of four hidden layers with dimensions 1024, 512, 256, and 128, respectively. Training data are generated in two stages: first by sampling UC decisions, followed by scenario-based sampling of net load profiles. To efficiently construct a large-scale dataset, multi-period optimal power flow (OPF) computations are executed in parallel. A total of 1 million samples are used for training the network. All benchmark results are obtained using Gurobi 12.0, running on a workstation with an Intel i9-9900K @ 3.60 GHz CPU and 64 GB RAM.

\vspace*{-1mm}
\subsection{Test Results}
\vspace*{-1mm}
\begin{table}[t]
\centering
\vspace{-3mm}
\caption{Performance Comparison on the 118-Bus System}
\vspace{-1mm}
\label{table:test_results}
\renewcommand{\arraystretch}{1.2}
\resizebox{\linewidth}{!}{%
\begin{tabularx}{1.1\linewidth}{llllll}
\toprule
\multirow{2}{*}{\vspace{-2mm}\textbf{Method}} & \multirow{2}{*}{\vspace{-2mm}\textbf{Metric}} & \multicolumn{4}{c}{\textbf{Number of Scenarios}} \\
\cmidrule(ll){3-6}
& & 100 & 200 & 500 & 1{,}000 \\
\midrule
\multirow{2}{*}{Gurobi}    
& Objective (\$)   & 1{,}615{,}052 & 1{,}614{,}822 & 1{,}614{,}218 & 1{,}613{,}558 \\
& Time (s)         & 4{,}340 & 8{,}632 & 22{,}562 & 46{,}310 \\
\midrule
\multirow{3}{*}{CCG}       
& Gap (\%)         & 0.036 & 0.056 & 0.062 & 0.068 \\
& Time (s)         & 1{,}350 & 2{,}256 & 5{,}232 & 9{,}213 \\
& Speedup          & \textbf{3.21$\times$} & \textbf{3.82$\times$} & \textbf{4.31$\times$} & \textbf{5.02$\times$} \\
\midrule
\multirow{3}{*}{Proposed} 
& Gap (\%)         & 0.058 & 0.072 & 0.083 & 0.096 \\
& Time (s)         & 251 & 283 & 332 & 356 \\
& Speedup          & \textbf{17.29$\times$} & \textbf{30.50$\times$} & \textbf{67.96$\times$} & \textbf{130.08$\times$} \\
\bottomrule
\end{tabularx}%
}
\vspace{-5mm}
\end{table}



Table~\ref{table:test_results} presents a comparative performance analysis of Gurobi, the classical CCG method, and the proposed neural CCG method on the IEEE 118-bus system across varying numbers of scenarios. As expected, Gurobi delivers the best objective values and serves as the benchmark for optimality, albeit with prohibitively long computation times. The CCG method substantially reduces runtime while maintaining high solution quality, with optimality gaps consistently below 0.07\%. The proposed neural CCG method achieves the highest computational efficiency, with speedups ranging from 17.29$\times$ to 130.08$\times$ relative to Gurobi. Although its optimality gaps are marginally higher, up to 0.096\%, they remain well within acceptable limits for practical applications. These results underscore a favorable trade-off between computational speed and solution accuracy, demonstrating that the proposed method is particularly well-suited for large-scale, scenario-based optimization tasks where rapid decision-making is critical.

\vspace{-1mm}
\section{Conclusion}
This letter proposes a Neural CCG method for efficiently solving large-scale 2S-SUC problems. By approximating the second-stage resource problem with a neural network, the proposed method significantly accelerates the solution process while maintaining solution quality. Numerical experiments on the IEEE 118-bus system demonstrate substantial speedups over the conventional CCG approach, with minimal loss in optimality. These results highlight the potential of integrating machine learning models into classical optimization frameworks to effectively address the scalability challenges of stochastic power system operations.

\vspace{-3mm}
\appendix
\subsection{Proof of Proposition 1} \label{appendix-proof}

\begin{proof}
Let $\mathcal{Z}$ be the finite set of first-stage decisions and $\mathcal{S}$ the finite set of all scenarios. At iteration $k$ the algorithm holds a scenario subset $\mathcal{S}^k \subseteq \mathcal{S}$ and a candidate decision $z_k \in \mathcal{Z}$.
Define $l_{k}(\bm{z}^*) := \max_{s \in \mathcal{S}} \mt{NN}(\bm{z}^*, \bm{\xi}_{s}),~~r_{k}(\bm{z}^*) := \max_{s \in \mathcal{S}^k} \mt{NN}(\bm{z}^*, \bm{\xi}_{s}) $. The algorithm stops when

\vspace{-5mm}
\begin{equation}
l_{k}(\bm{z}_k) - r_{k}(\bm{z}_k) \leq \varepsilon. \label{eq:terminate2}
\end{equation}
\vspace{-5mm}

Assume, to the contrary, that the algorithm generates an infinite sequence of iterations. Because $\mathcal{Z}$ is finite, there exists at least one decision $\bm{z}^* \in \mathcal{Z}$ selected infinitely often. Restrict attention to that infinite subsequence (indexed again by $k$ for simplicity). For every such $k$, \eqref{eq:terminate2} is violated, so

\vspace{-5mm}
\begin{equation}
l_{k}(\bm{z}^*) > r_{k}(\bm{z}^*) + \varepsilon. \label{eq:viol}
\end{equation}
\vspace{-5mm}




Because of \eqref{eq:viol}, the algorithm adds a scenario $\bm{\xi}_{s}^{*}$ to $\mathcal{S}^{k+1}$ satisfying $\mt{NN}(\bm{z}^*, \bm{\xi}_{s}^*) = l_{k}(\bm{z}^*)$. 

Hence $r_{k+1}(\bm{z}^*) = \max\{r_{k}(\bm{z}^*), l_{k}(\bm{z}^*) > r_{k}(\bm{z}^*) + \varepsilon \}$. Therefore, the sequence $\{r_{k}(\bm{z}^*)\}$ is strictly increasing by at least $\varepsilon$ at every iteration in the subsequence.


Because $\mathcal{Z}$ and $\mathcal{S}$ are finite and $\mt{NN} \left(\bm{z}, \bm{\xi}_{s} \right)$ is bounded on $\mathcal{Z} \times \mathcal{S}$, there exists $B$ such that:
$r_{k}(\bm{z}^*) < B, \forall k$.

An infinite sequence that increases by at least the fixed amount $\varepsilon > 0$ cannot remain bounded above by $B$. This contradiction shows that our initial assumption of infinitely many iterations is false. Consequently, algorithm \ref{alg:neu-ccg} must satisfy condition \eqref{eq:terminate} after a finite number of iterations and thus terminates.

\end{proof}

\vspace{-7mm}
\bibliographystyle{IEEEtran}
\bibliography{Refs}
\end{document}